

\input lanlmac

\input epsf

\newcount\figno
\figno=0
\def\fig#1#2#3{
\par\begingroup\parindent=0pt\leftskip=1cm\rightskip=1cm\parindent=0pt
\baselineskip=11pt
\global\advance\figno by 1
\midinsert
\epsfxsize=#3
\centerline{\epsfbox{#2}}
\vskip 12pt
\centerline{{\bf Fig.\the\figno~} #1}\par
\endinsert\endgroup\par
}
\def\figlabel#1{\xdef#1{\the\figno}}

\def\N{{\cal N}}

\def\th{\theta}

\def\cob{\delta}

\def\Tr{{\rm Tr}}
\def\hf{{1\over 2}}
\def\qu{{1\over 4}}
\def\R{{\bf R}}
\def\o{\over}
\def\til#1{\widetilde{#1}}

\def\del{\partial}

\def\bra{\langle}
\def\ket{\rangle}
\def\lf{\left}
\def\ri{\right}
\def\riya{\rightarrow}

\def\h#1{\widehat{#1}}

\def\bt{\beta}
\def\ga{\gamma}

\def\dag{\dagger}
\def\rt#1{\sqrt{#1}}
\def\st{\star}

\def\sitarel#1#2{\mathrel{\mathop{\kern0pt #1}\limits_{#2}}}

\def\tV{\widetilde{V}}
\def\tv{\widetilde{v}}

\def\K{{\rm K}_1}
\def\fHK{f^{{\rm HK}}}
\def\fmid{f^{{\rm mid}}}

\def\np#1#2#3{{ Nucl. Phys.} {\bf B#1} (#2) #3}

\def\plo#1#2#3{{ Phys. Lett.} {\bf #1B} (#2) #3}

\lref\HKcheck{
L.~Rastelli, A.~Sen and B.~Zwiebach,
``A note on a proposal for the tachyon state in vacuum string field  theory,''
[hep-th/0111153]\semi
R.~Rashkov and K.~S.~Viswanathan,
``A Note on the Tachyon State in Vacuum String Field Theory,''
[hep-th/0112202].
}
\lref\HI{
A.~Hashimoto and N.~Itzhaki,
``Observables of string field theory,''
[hep-th/0111092].
}
\lref\KishiOh{
I.~Kishimoto and K.~Ohmori,
``CFT description of identity string field: 
Toward derivation of the VSFT  action,''
[hep-th/0112169].
}
\lref\siegel{
K.~Okuyama,
``Siegel gauge in vacuum string field theory,''
[hep-th/0111087].
}
\lref\spec{
L.~Rastelli, A.~Sen and B.~Zwiebach,
``Star algebra spectroscopy,''
[hep-th/0111281].
}
\lref\GRSZ{
D.~Gaiotto, L.~Rastelli, A.~Sen and B.~Zwiebach,
``Ghost Structure and Closed Strings in Vacuum String Field Theory,''
[hep-th/0111129].
}

\lref\Witten{
E. Witten,
``Noncommutative Geometry And String Field Theory,''
Nucl.\ Phys.\  {\bf B268} (1986) 253.
}
\lref\HorowitzYZ{
G.~T.~Horowitz and A.~Strominger,
``Translations As Inner Derivations And Associativity Anomalies In Open
String Field Theory,''
Phys.\ Lett.\  {\bf B185} (1987) 45.
}
\lref\LPP{
A. LeClair, M. E. Peskin and C. R. Preitschopf,
``String Field Theory on the Conformal Plane. 1. Kinematical Principles,''
Nucl. Phys. {\bf B317} (1989) 411\semi
``String Field Theory on the Conformal Plane. 2. Generalized Gluing,''
Nucl. Phys. {\bf B317} (1989) 464.
}
\lref\Samuel{
S.~Samuel,
``The Physical and Ghost Vertices in Witten's String Field Theory,''
\plo{181}{1986}{255}.
}
\lref\GrossJ{
D. J. Gross and A. Jevicki,
``Operator Formulation of Interacting String Field Theory,''
\np{283}{1987}{1};
``Operator Formulation of Interacting String Field Theory (II),''
\np{287}{1987}{225}.
}
\lref\CST{
E.~Cremmer, A.~Schwimmer and C.~Thorn,
``The Vertex Function in Witten's Formulation of String Field Theory,''
\plo{179}{1986}{57}.
}
\lref\ItohWM{
K.~Itoh, K.~Ogawa and K.~Suehiro,
``BRS Invariance of Witten's Type Vertex,''
Nucl.\ Phys.\  {\bf B289} (1987) 127.
}

\lref\RZ{
L. Rastelli and B. Zwiebach,
``Tachyon potentials, star products and universality,''
JHEP {\bf 0109}, 038 (2001)
[hep-th/0006240].
}

\lref\VSFT{
L.~Rastelli, A.~Sen and B.~Zwiebach,
``String Field Theory Around the Tachyon Vacuum,''
[hep-th/0012251];
``Vacuum string field theory,''
[hep-th/0106010].
}

\lref\KP{
V. A. Kostelecky and R. Potting,
``Analytical construction of a nonperturbative vacuum
for the open bosonic string,''
Phys. Rev. {\bf D63} (2001) 046007 [hep-th/0008252].
}

\lref\Ohmori{
K.~Ohmori,
``A review on tachyon condensation in open string field theories,''
[hep-th/0102085].
}

\lref\RSZ{
L. Rastelli, A. Sen and B. Zwiebach,
``Half-strings, Projectors, and Multiple D-branes
in Vacuum String Field Theory,''
[hep-th/0105058];
``Boundary CFT construction of D-branes in vacuum string field theory,''
[hep-th/0105168];
}
\lref\RSZclas{
L. Rastelli, A. Sen and B. Zwiebach,
``Classical Solutions in String Field Theory Around the Tachyon Vacuum,''
[hep-th/0102112];
}
\lref\GT{
D. J. Gross and W. Taylor,
``Split string field theory I,'' JHEP {\bf 0108}, 009 (2001),
[hep-th/0105059];
``Split string field theory. II,''
JHEP {\bf 0108}, 010 (2001)
[hep-th/0106036].
}
\lref\KO{
T.~Kawano and K.~Okuyama,
``Open string fields as matrices,''
JHEP {\bf 0106}, 061 (2001)
[hep-th/0105129].
}
\lref\JD{
J.~R.~David,
``Excitations on wedge states and on the sliver,''
JHEP {\bf 0107}, 024 (2001)
[hep-th/0105184].
}
\lref\Mu{
P.~Mukhopadhyay,
``Oscillator representation of the BCFT construction of D-branes in  vacuum string field theory,''
[hep-th/0110136].
}
\lref\HK{
H.~Hata and T.~Kawano,
``Open string states around a classical 
solution in vacuum string field  theory,''
JHEP {\bf 0111}, 038 (2001)
[hep-th/0108150].
}
\lref\matsuo{
Y.~Matsuo,
``BCFT and sliver state,''
Phys.\ Lett.\ B {\bf 513}, 195 (2001)
[hep-th/0105175];
``Identity projector and D-brane in string field theory,''
Phys.\ Lett.\ B {\bf 514}, 407 (2001)
[hep-th/0106027];
``Projection operators and D-branes in purely 
cubic open string field  theory,''
Mod.\ Phys.\ Lett.\ A {\bf 16}, 1811 (2001)
[hep-th/0107007].
}
\lref\Kishimoto{
I.~Kishimoto,
``Some properties of string field algebra,''
JHEP {\bf 0112}, 007 (2001)
[hep-th/0110124].
}
\lref\HataMoriyama{
H.~Hata and S.~Moriyama,
``Observables as Twist Anomaly in Vacuum String Field Theory,''
[hep-th/0111034].
}
\lref\FO{
K.~Furuuchi and K.~Okuyama,
``Comma vertex and string field algebra,''
JHEP {\bf 0109}, 035 (2001)
[hep-th/0107101].
}
\lref\Moeller{
N.~Moeller,
``Some exact results on the matter star-product 
in the half-string  formalism,''
[hep-th/0110204].
}
\lref\moore{
G. Moore and W. Taylor,
``The singular geometry of the sliver,''
[hep-th/0111069].
}

\Title{             
                                             \vbox{\hbox{EFI-02-59}
                                             \hbox{hep-th/0201015}}}
{\vbox{
\centerline{Ghost Kinetic Operator of Vacuum String Field Theory}
}}

\vskip .2in

\centerline{Kazumi Okuyama}

\vskip .2in

\centerline{ Enrico Fermi Institute, University of Chicago} 
\centerline{ 5640 S. Ellis Ave., Chicago IL 60637, USA}
\centerline{\tt kazumi@theory.uchicago.edu}

\vskip 3cm
\noindent

Using the data of eigenvalues and eigenvectors of Neumann matrices
in the 3-string vertex,
we prove analytically  
that the ghost kinetic operator of vacuum string field theory
obtained by Hata and Kawano is equal to the ghost operator
inserted at the open string midpoint.  
We also comment on the values of
determinants appearing in the norm of sliver state.

\Date{January 2002}

\vfill
\vfill

\newsec{Introduction}
Vacuum string field theory (VSFT)
\refs{\VSFT} is proposed 
as the theory around the closed string vacuum after the 
open string tachyon condensation.
(See \refs{\RSZclas\RSZ\GT\KO\matsuo\JD\FO\HK\Kishimoto\Mu\Moeller
\HataMoriyama\moore\siegel\HI\GRSZ\HKcheck\spec{--}\KishiOh} 
for related papers.)
Since all the information of tachyon condensation is
contained in the kinetic operator $Q$ of this theory,
it is important to determine its form and study its properties.
One possible form of $Q$ is
\eqn\Qosci{
Q=c_0+\sum_{n=1}^{\infty}(c_n+(-1)^nc_{-n})f_n
}
where $f_n$'s are numerical coefficients. This satisfies the requirement
that $Q$ has a trivial cohomology and it is a derivation of the star product.

In \HK, Hata and Kawano found that the coefficients $f_n$ are 
determined uniquely if we demand that the equation of motion has a
nontrivial solution in the Siegel gauge.
The coefficients $\fHK_n$ they found are written 
in terms of the Neumann coefficients in the 3-string vertex of ghost
sector \refs{\HK,\Kishimoto}: 
\eqn\fHataKawa{
\fHK={1\o1-\til{M}}\til{v}.
}
See section 2 for the definition of $\til{M}$ and $\til{v}$.

Recently, Gaiotto, Rastelli, Sen and Zwiebach \GRSZ\ discussed 
that there is a canonical choice of kinetic operator:
the ghost insertion at the open string midpoint 
\eqn\Qmid{
Q=c\lf({\pi\o2}\ri)=\sum_{n=-\infty}^{\infty}c_n\cos{\pi n\o2}=c_0+
\sum_{n=1}^{\infty}(c_n+(-1)^nc_{-n})\cos{\pi n\o2}.
}
This operator has a form \Qosci\ with the coefficients $f_n$
given by
\eqn\deffnmid{
\fmid_n=\cos{\pi n\o2}=\hf\Big[i^n+(-i)^n\Big].
}
They conjectured that the operator found in \HK\ is 
nothing but this canonical kinetic operator, {\it i.e.}, 
\eqn\fHKeqfmid{
\fHK=\fmid,
}
and analyzed  this equality numerically.

In this paper, we will prove \fHKeqfmid\ analytically.
To do that, detailed information of the spectrum of
matrix $\til{M}$ is needed.
Since $\til{M}$ can be written in terms of the Neumann matrix $M$
in the matter sector,  
we can use the spectrum of $M$
recently obtained in \spec.

This paper is organized as follows: In section 2, we review the spectrum of 
Neumann matrices obtained in \spec. In section 3, we show that the eigenvector
of Neumann matrix is $\cob$-function normalizable. 
In section 4, we prove the equation
$\fHK=\fmid$ by showing that their generating functions coincide. 
In section 5, we define the index of Neumann matrix
and compute it for some examples. In section 6, we estimate the values
of determinants which appear in the norm of sliver state.
It is found that the norm of the matter part is vanishing
and the norm of the ghost part is divergent.

\newsec{Eigenvalues and Eigenvectors of Neumann Matrices}
In this section, we review the result of \spec. The 3-string vertex 
in the zero-momentum sector is given by \GrossJ
\eqn\Vthree{
|V_3\ket=\exp\lf(\sum_{r,s=1}^3 -\hf a^{(r)\dag}V^{rs}a^{(s)\dag}
+c^{(r)\dag}\til{V}^{rs}b^{(s)\dag}
+c^{(r)\dag}\til{v}^{rs}b_0^{(s)}\ri)
\bigotimes_{r=1}^3c_0^{(r)}c_1^{(r)}|0\ket_{r}.
}
In this paper, we use the following notation for Neumann matrices:  
\eqn\defMtMtv{
M=CV^{11},\quad \til{M}=C\tV^{11},\quad \tv=\tv^{11},
}
where $C_{nm}=(-1)^n\cob_{nm}$ is the twist matrix.
Note that $M$, $\til{M}$ and $\tv$ are twist-even:
\eqn\MCcom{
[M,C]=[\til{M},C]=0,\quad C\tv=\tv.
}

Since $|V_3\ket$ is invariant under the action of $L_1+L_{-1}$,
the matter Neumann matrix $M$ and the matrix $\K$ 
can be simultaneously diagonalized, 
where $\K$ is defined as the representation matrix
of $L_1+L_{-1}$ on matter oscillators:
\eqn\Konemat{
[L_1+L_{-1},v\cdot a]=(\K v)\cdot a-\rt{2}v_1p.
}
Here $v\cdot a=\sum_{n=1}^{\infty}v_na_n$.
The explicit form of $\K$ is given by
\eqn\compK{
(\K)_{nm}=-\rt{n(n-1)}\cob_{n-1,m}-\rt{n(n+1)}\cob_{n+1,m}.
}
Since $\K$ is a real and symmetric matrix, its eigenvalue is a real number.
The spectrum of $\K$ is continuous on the real axis.
The eigenvector $v^{(k)}$ of $\K$ with eigenvalue $k$
is implicitly given by the generating function
\eqn\genfunk{
f_k(z)=\sum_{n=1}^{\infty}{v^{(k)}_n\o\rt{n}}z^n=
{1\o k}(1-e^{-k\tan^{-1}z}).
}
This function has a symmetry 
\eqn\twistfk{
f_k(-z)=-f_{-k}(z),
}
which reflects the fact that $\K$ is twist-odd
\eqn\KC{
\{\K,C\}=0.
} 
From $f_{k=0}(z)=\tan^{-1}z$, 
the eigenvector $v^{(0)}$ with eigenvalue $k=0$ becomes
\eqn\vzero{
v^{(0)}_{2l}=0,\quad v^{(0)}_{2l-1}={(-1)^{l-1}\o\rt{2l-1}}.
}  

It is convenient to introduce a bracket notation 
for the infinite summation in \genfunk: 
\eqn\zEkfk{
f_k(z)=\bra z|E^{-1}|k\ket=\bra k|E^{-1}|z\ket,
}
where $E$ is a diagonal matrix defined by
\eqn\matE{
E_{nm}=\rt{n}\cob_{n,m},
}
and $|z\ket$ and $|k\ket$ denote the infinite-dimensional vectors 
\eqn\zkvec{
|z\ket=(z,z^2,z^3,\cdots)^T,\quad 
|k\ket=(v_1^{(k)},v^{(k)}_2,v^{(k)}_3,\cdots)^T.
}
$\bra z|=|z\ket^T$ is the transpose of $|z\ket$, 
not the hermitian conjugate of $|z\ket$.
Under the twist, $|z\ket$ and $|k\ket$ transform as
\eqn\Twistzk{
C|z\ket=|-z\ket,\quad C|k\ket=-|-k\ket.
}
A useful relation satisfied by $\bra z|$ is
\eqn\delbraz{
z\del_z\bra z|=\bra z|E^2.
}

For a matrix $X$ which commutes with $\K$, let $X(k)$ denote
the eigenvalue for the eigenvector $|k\ket$
\eqn\defXk{
X|k\ket=X(k)|k\ket.
}
The eigenvalue of $M$ is given by
\eqn\Mk{
M(k)=-{1\o1+2\cosh{\pi k\o2}}.
}
The width-matrix $T_N$ of the wedge state \RZ, 
which is defined by
\eqn\defwedge{
|N\ket_w=(|0\ket)_{\st}^{N-1}=\exp\lf(-\hf a^{\dag}CT_Na^{\dag}\ri)|0\ket,
}
also commutes with $\K$. $T_N$ can be written as \FO
\eqn\TNform{
T_N={T+(-T)^{N-1}\o1-(-T)^N},
}
where $T$ is the width-matrix of the sliver \refs{\KP,\RSZclas}
\eqn\Tform{
T={1\o2M}\lf(1+M-\rt{(1-M)(1+3M)}\ri).
}
Note that $T_3=M$ and $T_{\infty}=T$.
The eigenvalues of $T$ and $T_N$ are given by
\eqn\Tk{
T(k)=-e^{-{\pi\o2}|k|}, \quad
T_N(k)={\sinh\lf({2-N\o4}\pi k\ri)\o\sinh\lf({N\o4}\pi k\ri)}.
}

\newsec{Inner Product $\bra k|p\ket$}
In this section, we study the inner product between 
two eigenvectors $|k\ket,|p\ket$ of $\K$. 
The inner product of two vectors is defined by
\eqn\innerP{
\bra v|v'\ket\equiv\sum_{n=1}^{\infty}v_nv'_n
=\int_{-{\pi\o2}}^{{3\pi\o2}}{d\th\o2\pi}\bra v|e^{i\th}\ket
\bra e^{-i\th}|v'\ket.
}
Since $\K$ is symmetric, we can expect that
two eigenvectors with different eigenvalues
are orthogonal to each other with respect to this inner product.
But the norm of eigenvector is divergent, which can be seen from the example
\vzero. Therefore, the eigenvector of $\K$ is non-normalizable in this 
sense. However, as we will show below,
the eigenvector of $\K$ is $\cob$-function normalizable
as usual for the continuous spectrum.
To see this, 
let us calculate the inner product $\bra k|p\ket$ of two eigenvectors:
\eqn\innerkp{\eqalign{
\bra k|p\ket&=\int_{-{\pi\o2}}^{{3\pi\o2}}{d\th\o2\pi}
\bra k|E^{-1}|e^{i\th}\ket\bra e^{-i\th}|E|p\ket \cr
&=\int_{-{\pi\o2}}^{{3\pi\o2}}{d\th\o2\pi}
\bra k|E^{-1}|e^{i\th}\ket 
\Big(z\del_z\bra z|E^{-1}|p\ket\Big|_{z=e^{-i\th}}\Big) \cr
&=\int_{-{\pi\o2}}^{{3\pi\o2}}{d\th\o2\pi}
{1\o k}\Big(1-e^{-k\tan^{-1}e^{i\th}}\Big)
\Big({1\o2\cos\th}e^{-p\tan^{-1}e^{-i\th}}\Big)
}}
Here we used \delbraz, \zEkfk\ and \genfunk.
\fig{Image of the unit circle under the map $f(z)=\tan^{-1}z$}
{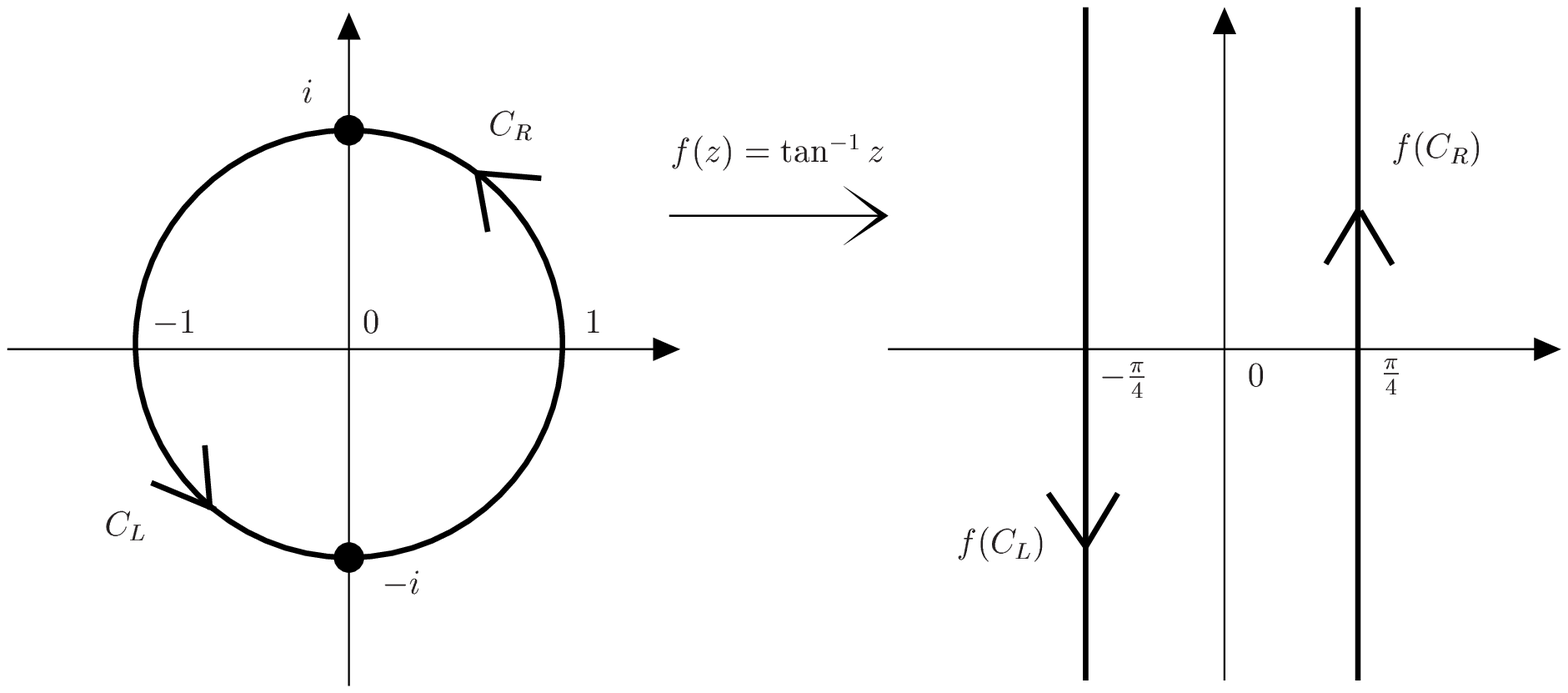}{130mm}
In order to evaluate this integral, we should specify the 
branch of the function
\eqn\arctan{
\tan^{-1}z={1\o2i}\log{1+iz\o1-iz}.
}
For the consistency of the relation $f_k(0)=0$ \genfunk, we should take
a branch such that $\tan^{-1}(0)=0$.
In other words, $\tan^{-1}z$ is given by the Taylor series
\eqn\arctanTay{
\tan^{-1}z=\sum_{l=1}^{\infty}{(-1)^{l-1}\o2l-1}z^{2l-1}.
} 
On this branch of the map $w=\tan^{-1}z$,
the unit disk $|z|<1$ is mapped to the strip 
$|{\rm Re}\,w|<\pi/4$, 
the right half of the unit circle $(C_R)$ is mapped to the line ${\pi\o4}+i\R$
on the $w$-plane, and the left half $(C_L)$ is mapped to $-{\pi\o4}+i\R$
(see Fig.1):
\eqn\brancharctan{\eqalign{
\tan^{-1}e^{i\th}&={\pi\o4}+{i\o2}\log\tan\lf({\th\o2}+{\pi\o4}\ri)\quad
{\rm on}~~C_R=\lf\{-{\pi\o2}\leq \th\leq{\pi\o2}\ri\}, \cr
&=-{\pi\o4}-{i\o2}\log\tan\lf({\th\o2}-{\pi\o4}\ri)\quad
{\rm on}~~C_L=\lf\{{\pi\o2}\leq\th\leq{3\pi\o2}\ri\}.
}}

Let us first consider the contribution from $C_R$. 
We make a change of integration variable from $\th$ to $x$ defined by 
\eqn\defxtan{
\tan^{-1}e^{i\th}={\pi\o4}+ix,\quad
\tan^{-1}e^{-i\th}={\pi\o4}-ix,\quad \tan{\th\o2}=\tanh x,
}
\eqn\measuredx{
{d\th\o2\cos\th}=dx,\quad d\th={2dx\o\cosh 2x}.
}
Note that $x$ runs from $-\infty$ to $\infty$. 
Then, the contribution from $C_R$ to the integral \innerkp\ 
becomes
\eqn\intCR{\eqalign{
\int_{C_R}{d\th\o2\pi}{1\o k}\Big(1-e^{-k\tan^{-1}e^{i\th}}\Big)
\Big({1\o2\cos\th}e^{-p\tan^{-1}e^{-i\th}}\Big) 
&=\int_{-\infty}^{\infty}{dx\o2\pi}{1\o k}(1-e^{-{\pi k\o4}-ikx})
e^{-{\pi p\o4}+ipx} \cr
&={1\o k}\cob(p)-{1\o k}e^{-{\pi k\o2}}\cob(k-p).
}}
Next we consider the integral over $C_L$. Here we also change the integration
variable as
\eqn\intvy{
\tan^{-1}e^{i\th}=-{\pi\o4}-ix,\quad
\tan^{-1}e^{-i\th}=-{\pi\o4}+ix,\quad -\cot{\th\o2}=\tanh x,
}
\eqn\measuredy{
{d\th\o2\cos\th}=-dx,\quad d\th={2dx\o\cosh 2x}.
}
The contribution from $C_L$ is given by
\eqn\intCL{\eqalign{
\int_{C_L}{d\th\o2\pi}{1\o k}\Big(1-e^{-k\tan^{-1}e^{i\th}}\Big)
\Big({1\o2\cos\th}e^{-p\tan^{-1}e^{-i\th}}\Big) 
&=-\int_{-\infty}^{\infty}{dx\o2\pi}{1\o k}(1-e^{{\pi k\o4}+ikx})
e^{{\pi p\o4}-ipx} \cr
&=-{1\o k}\cob(p)+{1\o k}e^{{\pi k\o2}}\cob(k-p).
}}
By adding \intCR\ and \intCL, we can see that $\bra k|p\ket$ is 
proportional to the $\cob$-function
\eqn\normkp{
\bra k|p\ket=\N(k)\cob(k-p),
}
where $\N(k)$ is given by
\eqn\Nkform{
\N(k)={2\o k}\sinh{\pi k\o2}.
}
From \intCR\ and \intCL, $k$ can be regarded as an analogue of momentum
along the lines $\pm{\pi\o4}+i\R$ on the $w$-plane.

We can introduce the normalized eigenvector $|\h{k}\ket$ of $\K$  by
\eqn\unitk{
|\h{k}\ket=\N(k)^{-\hf}|k\ket,
}
whose inner product is given by the $\cob$-function
\eqn\unitnorm{
\bra\h{k}|\h{p}\ket=\cob(k-p).
}

In terms of this basis, we can write down the completeness 
relation
\eqn\onekk{
{\bf 1}=\int_{-\infty}^{\infty}dk|\h{k}\ket\bra\h{k}|=
\int_{-\infty}^{\infty}dk \N(k)^{-1}|k\ket\bra k|.
}
As a consistency check of \onekk, let us prove the following relation
\eqn\zwcomp{
\bra z|w\ket=\int_{-\infty}^{\infty}dk\bra z|E|\h{k}\ket\bra\h{k}|E^{-1}|w\ket
}
for arbitrary points $z,w$ on the unit disk.
The left-hand-side is given by
\eqn\zwinn{
\bra z|w\ket=\sum_{n=1}^{\infty}z^nw^n={zw\o1-zw}.
}
The right-hand-side is calculated as 
\eqn\zwkrepinn{\eqalign{
&\int_{-\infty}^{\infty}dk\bra z|E|\h{k}\ket\bra\h{k}|E^{-1}|w\ket\cr
=&z\del_z\int_{-\infty}^{\infty}dk\bra z|E^{-1}|k\ket\N(k)^{-1}
\bra k|E^{-1}|w\ket \cr
=&{z\o1+z^2}\int_{-\infty}^{\infty}dk{k\o2\sinh{\pi k\o2}}e^{-k\tan^{-1}z}
{1\o k}(1-e^{-k\tan^{-1}w}) \cr
=&{z\o1+z^2}\int_0^{\infty}{dk\o\sinh{\pi k\o2}}\Big[
-\sinh (k\tan^{-1}z)+\sinh (k\tan^{-1}z+k\tan^{-1}w)\Big] \cr
=&{z\o1+z^2}\Big[-\tan(\tan^{-1}z)+\tan(\tan^{-1}z+\tan^{-1}w)\Big] \cr
=&{zw\o1-zw}.
}}
Here we have used the formula
\eqn\shformula{
\int_0^{\infty}dx{\sinh ax\o\sinh bx}={\pi\o2b}\tan{\pi a\o2b},\quad
(|{\rm Re}\,a|< {\rm Re}\, b).
}
Since the unit disk $|z|<1$ is mapped to the strip 
$|{\rm Re}(\tan^{-1}z)|<\pi/4$, the $k$-integral in \zwkrepinn\ converges.

\newsec{Proof of $\fHK=\fmid$}
In this section, we prove the equivalence of $\fHK$ and $\fmid$
using the result in the previous section. We will show that the generating
functions of $\fHK_n$ and $\fmid_n$ coincide:
\eqn\zfeq{
\bra z|\fHK\ket=\bra z|\fmid\ket.
}
The generating function of $\fmid$ is easily found to be
\eqn\zfmid{
\bra z|\fmid\ket=\sum_{n=1}^{\infty}z^n\cos{\pi n\o2}=
\sum_{l=1}^{\infty}(-1)^lz^{2l}=-{z^2\o1+z^2}.
}
We assume $|z|<1$ to make this summation converge.
In order to calculate $\bra z|\fHK\ket$, we have to know the
explicit form of Neumann coefficients in the ghost sector.
$\til{v}$ is given by \GrossJ
\eqn\tilvB{
\til{v}_{2l}={2\o3}B_{2l},\quad \til{v}_{2l-1}=0,
}
where $B_{2l}$ is defined by
\eqn\defBn{
\lf({1+iz\o1-iz}\ri)^{2\o3}=\exp\lf({4i\o3}\tan^{-1}z\ri)
=\sum_{n={\rm even}}B_nz^n+i\sum_{n={\rm odd}}B_nz^n.
}
The generating function of $\tv$ becomes
\eqn\ztvform{
\bra z|\tv\ket={2\o3}\cosh\lf({4\o3}i\tan^{-1}z\ri).
} 

The Neumann matrix $\til{M}$ in the ghost sector is 
related to the corresponding
matrix $M$ in the matter sector as \GrossJ  
\eqn\tilMM{
\til{M}=-E\lf({M\o1+2M}\ri)E^{-1}.
}
Thus, $\fHK$ can be written as
\eqn\fninAv{
\fHK={1\o1-\til{M}}\tv=EAE^{-1}\tv,
}
where
\eqn\defmatA{
A={1+2M\o1+3M}.
}
Therefore, we can use the information of the spectrum of $M$ to calculate
the generating function $\bra z|\fHK\ket$.
Plugging $M(k)$ of the form \Mk\ into \defmatA, the eigenvalue of
$A$ is found to be 
\eqn\eigenAk{
A(k)={2\cosh{\pi k\o2}-1\o2\cosh{\pi k\o2}-2}.
}

Using the completeness condition \onekk,
$\bra z|\fHK\ket$ can be rewritten as
\eqn\zfHKkrep{\eqalign{
\bra z|\fHK\ket&=\int_{-\infty}^{\infty}dk
\bra z|E|\h{k}\ket\bra\h{k}|E^{-1}|\fHK\ket \cr
&=z\del_z\int_{-\infty}^{\infty}dk 
\bra z|E^{-1}|k\ket \N(k)^{-1}\bra k|E^{-1}|\fHK\ket \cr
&={z\o1+z^2}\int_{-\infty}^{\infty}dk{e^{-k\tan^{-1}z}\o2\sinh{\pi k\o2}}
k\bra k|E^{-1}|\fHK\ket \cr
&=-{z\o1+z^2}\int_0^{\infty}dk{\sinh(k\tan^{-1}z)\o\sinh{\pi k\o2}}
k\bra k|E^{-1}|\fHK\ket.
}}
In the last step, we used the fact that 
$\bra k|E^{-1}|\fHK\ket$ is an odd-function of $k$ which follows from 
$C|k\ket=-|-k\ket$ and $C|\fHK\ket=|\fHK\ket$:
\eqn\oddtvint{
\bra -k|E^{-1}|\fHK\ket=-\bra k|CE^{-1}|\fHK\ket=-\bra k|E^{-1}C|\fHK\ket
=-\bra k|E^{-1}|\fHK\ket.
}

From \fninAv, the last factor in \zfHKkrep\ is written as
\eqn\kfHKasktv{
k\bra k|E^{-1}|\fHK\ket=A(k)k\bra k|E^{-1}|\tv\ket.
}
Therefore, the problem is reduced to the calculation 
of $k\bra k|E^{-1}|\tv\ket$. 
This quantity can be extracted from the generating function of $\tv$ \ztvform: 
\eqn\kinntvev{\eqalign{
k\bra k|E^{-1}|\tv\ket
&=\int_{-{\pi\o2}}^{{3\pi\o2}}{d\th\o2\pi}
k\bra k|E^{-1}|e^{i\th}\ket\bra e^{-i\th}|\tv\ket\cr
&=\int_{-{\pi\o2}}^{{3\pi\o2}}{d\th\o2\pi}(1-e^{-k\tan^{-1}e^{i\th}}){2\o3}
\cosh\lf({4\o3}i\tan^{-1}e^{-i\th}\ri).
}}
By changing the integration variable
as in the previous section, this integral can be written as
\eqn\kinntvint{\eqalign{
k\bra k|E^{-1}|\tv\ket&={8\o3}\int_{-\infty}^{\infty}{dx\o2\pi}{1\o\cosh 2x}
\lf[1-\cosh\lf({\pi k\o4}+ikx\ri)\ri]\cosh\lf({4\o3}x+{\pi i\o3}\ri)\cr
&={8\o3}\int_0^{\infty}{dx\o2\pi}{1\o\cosh 2x}\lf[
\cosh \lf({4\o3}x\ri)-\cosh\lf({4\o3}x+ikx\ri)\cosh\lf({\pi k\o4}+{\pi i\o3}\ri)\ri.\cr
&\hskip 51mm\lf.-\cosh\lf({4\o3}x-ikx\ri)
\cosh\lf({\pi k\o4}-{\pi i\o3}\ri)\ri].
}}
By making use of the formula
\eqn\coshform{
\int_0^{\infty}dx{\cosh ax\o \cosh bx}={\pi\o2b}\cdot{1\o\cos{\pi a\o2b}},\quad
(|{\rm Re}\,a|<{\rm Re}\,b),
}
the integral in \kinntvint\ is evaluated as
\eqn\kinntveval{
k\bra k|E^{-1}|\tv\ket={1\o3}\lf[2
-{\cosh\lf({\pi k\o4}+{\pi i\o3}\ri)\o\cosh\lf({\pi k\o4}-{\pi i\o3}\ri)}
-{\cosh\lf({\pi k\o4}-{\pi i\o3}\ri)\o\cosh\lf({\pi k\o4}+{\pi i\o3}\ri)}\ri]
={2\cosh{\pi k\o2}-2\o2\cosh{\pi k\o2}-1}={1\o A(k)}.
}
Therefore,
\eqn\kAkktvone{
k\bra k|E^{-1}|\fHK\ket=kA(k)\bra k|E^{-1}|\tv\ket=1.
}
Finally, \zfHKkrep\ becomes
\eqn\zfHKfinal{\eqalign{
\bra z|\fHK\ket&=-{z\o1+z^2}\int_0^{\infty}dk
{\sinh(k\tan^{-1}z)\o\sinh{\pi k\o2}}\cr
&=-{z\o1+z^2}\tan(\tan^{-1}z)=-{z^2\o1+z^2}=\bra z|\fmid\ket.
}}
Here we used the formula \shformula. This is valid since 
$|{\rm Re}(\tan^{-1}z)|<\pi/4$ for $|z|<1$.
This completes the proof of \zfeq.

We comment on the subtlety of \kAkktvone. Naively, one might
think that the following equation holds:
\eqn\naivefmid{
k\bra k|E^{-1}|\fmid\ket=k\bra k|E^{-1}|\fHK\ket.
}
However, the left-hand-side is ill-defined since $\tan^{-1}(i)=i\infty$:
\eqn\naivefmidk{
k\bra k|E^{-1}|\fmid\ket={k\o2}\Big[ f_k(i)+ f_k(-i)\Big]
=1-\cosh\Big(k\tan^{-1}(i)\Big).
}
By regularizing the last term of  \naivefmidk\
with a parameter $a(<1)$
\eqn\regfmidk{
k\bra k|E^{-1}|\fmid\ket=\lim_{a\riya 1}{k\o2}\Big[ f_k(ia)+ f_k(-ia)\Big]
=1-\lim_{a\riya 1}\cosh\Big(k\tan^{-1}(ia)\Big),
}
we can show that this term does not contribute 
to the computation of $\bra z|f\ket$:
\eqn\secondcont{\eqalign{
&\lim_{a\riya 1}\int_0^{\infty}dk
{\sinh(k\tan^{-1}z)\o\sinh{\pi k\o2}}\cosh\Big(k\tan^{-1}(ia)\Big) \cr
=&\lim_{a\riya 1}\hf\int_0^{\infty}dk
{1\o\sinh{\pi k\o2}}\lf[\sinh \Big(k\tan^{-1}z+k\tan^{-1}(ia)\Big)+
\sinh \Big(k\tan^{-1}z-k\tan^{-1}(ia)\Big)\ri] \cr
=&\lim_{a\riya 1}\hf\lf[{z+ia\o1-iaz}+{z-ia\o1+iaz}\ri]
=\lim_{a\riya 1}{(1-a^2)z\o1+a^2z^2}=0.
}}
Therefore, \naivefmid\ should be understood as an equation 
up to such an irrelevant term. Here we emphasize that
$\bra z|\fmid\ket$ is well-defined and is equal to $\bra z|\fHK\ket$
without any ambiguity.

\newsec{Index of Neumann Matrix}
As an application of our formalism, we define the index of 
Neumann matrix and compute it for some examples.
For a general matrix $X$ diagonalized on the basis $\{|k\ket\}$
with its eigenvalue $X(k)=X(-k)$ an even-function of $k$, 
the eigenvalue $X(k)=X(-k)$ has a two-fold degeneracy
\eqn\twisteigen{\eqalign{
|k,+\ket&=|k\ket+C|k\ket=|k\ket-|-k\ket, \cr
|k,-\ket&=|k\ket-C|k\ket=|k\ket+|-k\ket,
}}
where $|k,+\ket\,(~|k,-\ket~)$ 
is a twist-even (odd) eigenvector.
Note that $X$ is twist-even $[X,C]=0$ when 
$X(k)=X(-k)$.
For $k=0$, there is only one twist-odd eigenvector.
Therefore, the index of $X$ defined by   
\eqn\defInd{
I(X)=\Tr(CX)
}
receives contribution only from $k=0$.  
This can be thought of as a measure of the twist anomaly \HataMoriyama.
$I(X)$ can be calculated as
\eqn\IndXkzero{\eqalign{
I(X)&=\int_{-\infty}^{\infty}dk\bra\h{k}|CX|\h{k}\ket
=\int_{-\infty}^{\infty}dk\bra\h{k}|C|\h{k}\ket X(k)\cr
&=-\int_{-\infty}^{\infty}dk\bra\h{k}|-\h{k}\ket X(k)
=-\int_{-\infty}^{\infty}dk\cob(2k)X(k)=-\hf X(0).
}}
As expected, it depends only on the eigenvalue at $k=0$. 

For example, the index of $T_N$ is given by
\eqn\IndTN{
I(T_N)=-\hf T_N(0)=\hf-{1\o N}.
}
Here we used \Tk.
In the rest of this section, we check this prediction for some examples 
by directly computing 
the trace. 
We first consider  the index of $B$ defined by \spec
\eqn\defB{
B={d\o dN}T_N\Big|_{N=2}=-{T\log(-T)\o1-T^2}.
}
The eigenvalue of $B$ is given by 
\eqn\Bk{
B(k)=-\hf{{\pi k\o2}\o\sinh{\pi k\o2}}.
}
It is known that this matrix can be written explicitly 
in the $n,m$ basis \spec
\eqn\Bnm{
B_{nm}=-{1+(-1)^{n+m}\o2}{(-1)^{n-m\o2}\rt{nm}\o(n+m)^2-1}.
}
Therefore, we can compute the index of $B$ directly as
\eqn\TrCB{
I(B)=\sum_{n=1}^{\infty}(-1)^nB_{nn}=
-\sum_{n=1}^{\infty}(-1)^n{n\o (2n)^2-1}=
\qu\sum_{n=1}^{\infty}\lf({(-1)^{n-1}\o 2n-1}-{(-1)^n\o 2n+1}\ri)=\qu.
}
This agrees with the expected result
\eqn\IndBexp{
I(B)=-\hf B(0)={d\o dN}I(T_N)\Big|_{N=2}=\qu.
}

Next we consider the index of $T_3=M$. The diagonal element of 
$M$ is given by \GrossJ
\eqn\diagM{
M_{nn}=-{1\o3}\lf[2\sum_{k=0}^n(-1)^kA_k^2-1-(-1)^nA_n^2\ri]
}
where $A_k$ is defined by
\eqn\defukcoef{
\sum_{k={\rm even}}A_kz^k+i\sum_{k={\rm odd}}A_kz^k
=\lf({1+iz\o1-iz}\ri)^{1\o3}
=\exp\lf({2i\o3}\tan^{-1}z\ri).
}
The partial sum of the series $(-1)^nM_{nn}$ turns out to be
\eqn\partsumM{\eqalign{
\sum_{n=1}^L(-1)^nM_{nn}&={1\o3}\lf(1-\sum_{k=0}^L(-1)^kA_k^2\ri)
\quad L={\rm even}, \cr
&={1\o3}\sum_{k=0}^L(-1)^kA_k^2
\hskip 16mm L={\rm odd}.
}}
This sum converges in the limit $L\riya\infty$ since 
$\sum_{k=0}^{\infty}(-1)^kA_k^2=\hf$.
This can be shown by integrating the generating function of $A_k$ 
\defukcoef:
\eqn\Akintsum{\eqalign{
\sum_{k=0}^{\infty}(-1)^kA_k^2&=\int_{-{\pi\o2}}^{{3\pi\o2}}{d\th\o2\pi}
e^{{2i\o3}\tan^{-1}e^{i\th}}e^{{2i\o3}\tan^{-1}e^{-i\th}}\cr
&=\int_{C_R}{d\th\o2\pi}e^{{\pi i\o3}}+\int_{C_L}{d\th\o2\pi}e^{-{\pi i\o3}}
=\hf.
}}
Therefore, the index of $M$ computed in this way
agrees with the expected result \IndTN
\eqn\IndM{
I(M)=\sum_{n=1}^{\infty}(-1)^nM_{nn}={1\o6}.
}
It will be interesting to check \IndTN\ for general $N$ by directly
computing the trace.

We comment on a deformation of the index. We can introduce a parameter
$\bt$ in the definition of index, {\it e.g.},
\eqn\Ibt{
I_{\bt}(X)=\Tr(CXe^{-\bt\K^2}).
}
Naively, $I_{\bt}(X)$ is independent of $\bt$.
However, it can be $\bt$-dependent as in the case of
Witten index of a theory with continuous spectrum.
It is interesting to study the $\bt$-(in)dependence of $I_{\bt}(X)$.  

\newsec{Norm of Sliver}

\subsec{Eigenvalue Density $\rho(k)$}
In contrast with the finiteness of the index \defInd, 
the trace of $X$ itself is
divergent in general. This can be seen as
\eqn\Xdiag{
\Tr X=\int_{-\infty}^{\infty}dk \bra\h{k}|X|\h{k}\ket=
\int_{-\infty}^{\infty}dk \bra\h{k}|\h{k}\ket X(k)=
\int_{-\infty}^{\infty}dk \cob(0)X(k).
}
If we formally define the eigenvalue density $\rho(k)$ by
\eqn\defrhok{
\Tr X=\int_{-\infty}^{\infty}dk\rho(k)X(k),
}
then $\rho(k)$ is independent of $k$ and divergent 
\eqn\rhocob{
\rho(k)=\bra\h{k}|\h{k}\ket={\bra k|k\ket\o\N(k)}=\cob(0).
}
In the level truncation approximation, we can estimate the order of
divergence of $\rho(k)=\rho(0)$ as follows:
\eqn\rholeveltranc{
\rho(k)={\bra k=0|k=0\ket\o\N(0)}
\sim{1\o\N(0)}\sum_{n=1}^L\Big(v_n^{(0)}\Big)^2
\sim {1\o\pi}\sum_{l=1}^{L/2}{1\o2l-1}\sim {1\o2\pi}\log L\quad 
({\rm for~large}~L).
}
This expression agrees with the one obtained in \spec.
We can also estimate $\rho(k)$ by computing the trace of $B$
in two different ways:
\eqn\TrB{
\Tr B=\int_{-\infty}^{\infty}dk\rho(k)B(k)\sim \sum_{n=1}^LB_{nn}
}
where
\eqn\sumBintB{\eqalign{
\int_{-\infty}^{\infty}dk\rho(k)B(k)&=\rho(k)
\int_{-\infty}^{\infty}dk-\hf{{\pi k\o2}\o\sinh {\pi k\o2}}
=-{\pi\o2}\rho(k),\cr
\sum_{n=1}^{L}B_{nn}&=-\sum_{n=1}^{L}{n\o 4n^2-1}
\sim -\qu\log L.
}}
This again reproduce  \rholeveltranc.

Now it is natural to define the regularized trace $\Tr_L$ by
\eqn\defTrL{
\Tr_LX=\log L\int_{-\infty}^{\infty}{dk\o2\pi}X(k).
}
Using this trace, the determinant of $X$ can be regularized as
\eqn\regdet{
\det X\sim \exp\lf(\Tr_L\log X\ri)=L^{\ga(X)},
}
where the exponent $\ga(X)$ is defined by
\eqn\defgaX{
\ga(X)=\int_{-\infty}^{\infty}{dk\o2\pi}\log X(k).
}
When the eigenvalue $X(k)$ is an even-function of $k$,
$\ga(X)$ can be written as
\eqn\deteven{
\ga(X)= \int_0^{\infty}{dk\o\pi}\log X(k)
={2\o\pi^2}\int_0^1{du\o u}\log X(k)\Big|_{u=e^{-{\pi \o2}k}}.
} 

\subsec{Classical Solution of VSFT}
In this subsection, we review the classical solution $\Psi_0$ of VSFT
found in \HK.
$\Psi_0$ is given by
\eqn\solpsi{
|\Psi_0\ket=-\N_m\N_g\exp\lf(-\hf a^{\dag}CTa^{\dag}
+c^{\dag}C\til{T}b^{\dag}\ri)c_1|0\ket.
}
Here $T$ is defined by \Tform\ and $\til{T}$ is defined by
\eqn\deftilT{
\til{T}={1\o2\til{M}}\lf(1+\til{M}-\rt{(1-\til{M})(1+3\til{M})}\ri)=-ETE^{-1}.
}
The matter part of $\Psi_0$ is the sliver state and the ghost part 
is conjectured to be the twisted sliver state \GRSZ.  
$\N_m$ and $\N_g$ are the normalization constants given by
\eqn\normali{\eqalign{
\N_m&=\det{}^{{D\o2}}(1-M)(1+T), \cr
\N_g&=\det{}^{-1}(1-\til{M})(1+\til{T}),
}}
where $D(=26)$ is the dimension of D-brane.
The energy density of this solution is proportional to
\eqn\energysol{
{(2\pi)^D\o V_D}\bra\Psi_0|Q|\Psi_0\ket=E_mE_g,
}
where $V_D$ is the volume of D-brane and 
\eqn\EnEg{\eqalign{
E_m&=\det(1-M)^{{3\o4}D}(1+3M)^{{1\o4}D},\cr
E_g&=\det(1-\til{M})^{-{3\o2}}\det(1+3\til{M})^{-\hf}.
}}

\subsec{Calculation of Various Determinants}
We can estimate the values of determinants in \normali\ and \EnEg\
by using the regularization \regdet.
Let us compute the exponent $\ga$  \deteven\
for various matrices. 
One can show that $\ga$'s of Neumann matrices are written as 
combinations of the following integrals:
\eqn\Ivar{\eqalign{
I_-&=\int_0^1{du\o u}\log(1-u)=-\int_0^1{du\o u}
\sum_{n=1}^{\infty}{1\o n}u^n=-S_0=-{\pi^2\o6}, \cr
I_+&=\int_0^1{du\o u}\log(1+u)=\int_0^1{du\o u}
\sum_{n=1}^{\infty}{(-1)^{n-1}\o n}u^n
=-S_1={\pi^2\o12}, \cr
J_-&=\int_0^1{du\o u}\log(1-u+u^2)=\int_0^1{du\o u}
\log(1-e^{{\pi \o3}i} u)(1-e^{-{\pi \o3}i}u) \cr
&=-\int_0^1{du\o u}\sum_{n=1}^{\infty}{1\o n}u^n
(e^{{\pi n\o3}i}+e^{-{\pi n \o3}i})=-2S_{{1\o3}}=-{\pi^2\o18}, \cr
J_+&=\int_0^1{du\o u}\log(1+u+u^2) =\int_0^1{du\o u}
\log(1-e^{{2\pi \o3}i} u)(1-e^{-{2\pi \o3}i}u) \cr
&=-\int_0^1{du\o u}\sum_{n=1}^{\infty}{1\o n}u^n
(e^{{2\pi n \o3}i}+e^{-{2\pi n \o3}i})
=-2S_{{2\o3}}={\pi^2\o9}, 
}}
where $S_a$ is defined by
\eqn\defIa{
S_a\equiv 
\sum_{n=1}^{\infty}{1\o n^2}\cos(\pi na)={\pi^2\o4}(1-a)^2-{\pi^2\o12},
\quad (0\leq a\leq 2).
}
Substituting the eigenvalues \Mk\ and \Tk\ into the definition 
of $\ga$ \deteven, the exponents 
of the matter Neumann matrices are found to be
\eqn\variousga{\eqalign{
\ga(1+T)&={2\o\pi^2}I_- =-{1\o3},\hskip 15mm
\ga(1-T)={2\o\pi^2}I_+ ={1\o6} \cr
\ga(1-M)&={2\o\pi^2}(2I_+-J_+) ={1\o9},\hskip 5mm 
\ga(1+3M)={2\o\pi^2}(2I_--J_+) =-{8\o9} \cr
\ga(1+2M)&={2\o\pi^2}(J_--J_+) =-{1\o3},\quad 
\ga(1-TM)= {2\o\pi^2}(I_+-J_+) =-{1\o18}.
}}

For the ghost part, using the relation 
\eqn\ghMinmatM{
\til{T}=-ETE^{-1},\quad
1-\til{M}=E\lf({1+3M\o1+2M}\ri)E^{-1},\quad
1+3\til{M}=E\lf({1-M\o1+2M}\ri)E^{-1},
}
and assuming the property
\eqn\simdet{
\det(EXE^{-1})=\det X,
}
the exponents can be computed as
\eqn\gaghost{\eqalign{
\ga(1+\til{T})&=\ga(1-T)={1\o6}, \cr
\ga(1-\til{M})&=\ga(1+3M)-\ga(1+2M)=-{5\o9},\cr
\ga(1+3\til{M})&=\ga(1-M)-\ga(1+2M)={4\o9}.
}}
Combining these relations, the order of determinants \normali\ and \EnEg\
can be estimated:
\eqn\detest{\eqalign{
\N_m&\sim L^{-{1\o9}D},\quad \N_g\sim L^{{7\o18}}, \cr
E_m&\sim L^{-{5\o36}D},\quad E_g\sim L^{{11\o18}}.
}}

In the limit $L\riya \infty$, $\N_m$ and $E_m$ tend to vanish.
This behavior of the norm in the matter sector
was observed in the numerical calculation \RSZclas.  
On the other hand, the norm in the ghost sector diverges
in the limit $L\riya\infty$. In the early days of the study of VSFT,
it was expected that the divergent factor coming
from the ghost sector compensates the vanishing factor from the 
matter sector and in total the norm becomes finite \RSZclas.
However, our result \detest\ shows that the contribution from
the ghost sector is not large enough to compensate the matter contribution.
The total norm is still vanishing in the limit $L\riya\infty$.  

Recent discussion in \GRSZ\ is that the action 
with the kinetic operator \Qosci\
is a singular description of the theory around the
closed string vacuum and there is an infinite 
overall factor in front of the action due to a singular field redefinition.
Therefore, this infinite factor would cancel the 
vanishing norm of the sliver as discussed in \GRSZ.

\vskip 20mm
\centerline{{\bf Acknowledgement}}
I would like to thank Kentaro Hori and Sunny Itzhaki
for discussion and encouragement.

\listrefs

\end